\documentclass[twocolumn,trackchanges]{aastex63}
\usepackage{amsmath}

\newcommand{\bs}{\boldsymbol}
\newcommand{\nue}{\nu_{\rm e}}
\newcommand{\nueb}{\bar{\nu}_{\rm e}}
\newcommand{\nux}{\nu_x}
\newcommand{\ye}{Y_{\rm e}}

\submitjournal{ApJL}

\shorttitle{FFC in rotating CCSN}
\shortauthors{Harada et al.}
\graphicspath{{./}{figures/}}
\begin{document}

\title{Prospects of fast flavor neutrino conversion in rotating core-collapse supernovae}

\correspondingauthor{Akira Harada}
\author[0000-0003-1409-0695]{Akira Harada}
\affil{Interdisciplinary Theoretical and Mathematical Sciences Program (iTHEMS), RIKEN, Wako, Saitama 351-0198, Japan}
\author[0000-0002-7205-6367]{Hiroki Nagakura}
\affil{Division of Science, National Astronomical Observatory of Japan, 2-21-1 Osawa, Mitaka, Tokyo 181-8588, Japan}



\begin{abstract}
There is mounting evidence that neutrinos undergo fast flavor conversion (FFC) in core-collapse supernova (CCSN). In this paper, we investigate the roles of stellar rotation on the occurrence of FFC by carrying out axisymmetric CCSN simulations with full Boltzmann neutrino transport. Our result suggests that electron neutrino lepton number (ELN) angular crossings, which are the necessary and sufficient condition to trigger FFC, preferably occur in the equatorial region for rotating CCSNe. By scrutinizing the neutrino--matter interaction and neutrino radiation field, we find some pieces of evidence that the stellar rotation facilitates the occurrence of FFC. The low-electron-fraction region in the post-shock layer expands by centrifugal force, enhancing the disparity of neutrino absorption between electron-type neutrinos ($\nue$) and their anti-particles ($\nueb$). This has a significant impact on the angular distribution of neutrinos in momentum space, in which $\nue$ tends to be more isotropic than $\nueb$; consequently, ELN crossings emerge. The ELN crossing found in this study is clearly associated with rotation, which motivates further investigation on how the subsequent FFC influences explosion dynamics, nucleosynthesis, and neutrino signals in rotating CCSNe.
\end{abstract}



\section{Introduction} \label{sec:intro}

Core-collapse supernovae (CCSNe) are massive stellar explosions that announce the death of stars. During the last decade, remarkable progress has been made in the theory of CCSNe, which is greatly attributed to efforts on both multi-dimensional (multi-D) numerical modelings and improvements of input physics. On the other hand, there is a growing interest in collective neutrino oscillation in CCSNe theory. If it happens, the flavor conversion results in mixing the energy spectrum and angular distributions of each species of neutrinos, indicating that this potentially affects CCSN dynamics.

Fast flavor conversion (FFC), one of the modes of collective neutrino oscillation \citep{2005PhRvD..72d5003S, 2016PhRvL.116h1101S, 2016JCAP...03..042C}, has been of great interest recently. This mode offers the fastest flavor conversion in dense neutrino gases such as CCSN cores. The quantum kinetic equation (QKE) provides a way to investigate the property of FFC \citep{2015IJMPE..2441009V, 2016PhRvD..94c3009B}. However, the consistent treatment of neutrino flavor conversion, transport, and collision term is a formidable technical difficulty. Nevertheless, much effort has been expended in the study of non-linear features \citep{2020arXiv201101948T, 2019PhRvD.100b3016M, 2020PhLB..80035088M, 2020PhRvD.101b3007M, 2021PhRvD.103f3001M, 2021PhRvD.103h3013R, 2021arXiv210908631R, 2021arXiv210806356K, 2021arXiv210914011S, 2021PhRvL.126f1302B, 2021arXiv210900091S, 2019PhRvL.122i1101C, 2021arXiv211002286D, 2021arXiv210809886W, 2020PhRvD.102j3017J, 2021arXiv210410532Z}. On the other hand, linear stability analyses of flavor conversion, which can be carried out based on neutrino data from classical neutrino transport, provide an indication of whether FFC occurs in neutrino radiation fields \citep{2017PhRvL.118b1101I}. Very recently, \cite{2021arXiv210315267M} proved that a necessary and sufficient condition for the instability is the electron neutrino lepton number (ELN) crossing, which means that the angular distributions of electron-type neutrinos ($\nue$) and their anti-particles ($\nueb$) are crossing each other in momentum space. According to the recent studies of searching ELN crossings or linear stability analysis, FFC commonly occurs in the CCSN environment \citep{2019PhRvD.100d3004A, 2020PhRvD.101d3016A, 2020PhRvD.101b3018D, 2019ApJ...886..139N, 2020PhRvR...2a2046M, 2020PhRvD.101f3001G,2021arXiv210807281N, 2021PhRvD.103f3033A, 2020arXiv201208525C}, which is why FFC gains increased attention from the community.

It should be noted, however, that the linear stability analysis or ELN crossing search thus far has been focused only on CCSN models without stellar rotation. For rotational CCSNe, on the other hand, there are some characteristic properties of rotation in both fluid dynamics and neutrino radiation fields \citep{2018ApJ...852...28S, 2019ApJ...872..181H, 2020MNRAS.494.4665P, 2021MNRAS.508..966T}. By scrutinizing neutrino radiation fields computed from axisymmetric CCSN simulations with full Boltzmann neutrino transport, we investigate how the rotation gives impacts on the occurrence of FFC.

This paper is organized as follows. In Section~\ref{sec:method}, we provide the essence of our CCSN models and FFC analysis. Section~\ref{sec:evolution} provides information related to the supernova properties. In Section~\ref{sec:results}, we encapsulate our main findings. In Section~\ref{sec:concl}, we conclude and discuss the significance of our findings in the CCSN theory.

\section{Numerical setups} \label{sec:method}
The axisymmetric CCSN simulations presented in this paper are performed by a multi-D neutrino-radiation-hydrodynamics code. We simultaneously solve the Boltzmann equation for neutrino transport, the Euler equation for fluid, and the Poisson equation for gravity. The details of the code and their basic equations are described in \cite{2012ApJS..199...17S, 2014ApJS..214...16N, 2017ApJS..229...42N, 2019ApJ...878..160N} and references therein. 

In this study, we employ a progenitor model of $15\,M_\odot$ evolved without rotation presented in \cite{2002RvMP...74.1015W}. We run two CCSN simulations (with and without rotation) and start the simulations from the iron-collapse phase. For the rotating model, we add the rotational velocity of
\begin{equation}
v^\phi(r) = \frac{4\,{\rm rad\,s^{-1}}}{1+(r/10^3\,{\rm km})^2}
\end{equation}
by hand at the onset of the collapse, where $r$ is the radial coordinate measured from the grid center, not from the symmetry axis. We employ a multi-nuclear equation of state based on the variational method \citep{2017NuPhA.961...78T}. The neutrino reactions are taken from the standard set described in \citet{1985ApJS...58..771B} with some extensions \citep{2019ApJS..240...38N}. We consider three species of neutrinos: $\nue$, $\nueb$, and the other heavy-lepton-type neutrinos ($\nu_x$). Following the assumption, we take the two-flavor approximation in FFC analysis. We refer readers to see \citet{2020arXiv201208525C, 2020PhRvL.125y1801C} for cases distinguishing $\mu$ and $\tau$ neutrinos.
Throughout the simulation, we divide the radial range of $0\le r \le 5000\,{\rm km}$, neutrino energy of $0\le \epsilon \le 300\,{\rm MeV}$, and neutrino momentum angle of $0 \le \theta_\nu \le \pi$ and $0 \le \phi_\nu \le 2\pi$ into $384$, $20$, $10$, and $6$ grid points, respectively. The meridian plain $0 \le \theta \le \pi$ is divided into $64$ and $128$ grid points until and after the core bounce, respectively. At the core bounce, we imposed a random velocity perturbation of $0.1\%$.

To catch the essence of the roles of rotation on FFC, we focus on a representative time snapshot ($165\,{\rm ms}$ after bounce) and analyze the neutrino data by comparing the non-rotating and rotating models. We also estimate the linear growth rate of FFC by following an empirical formula given in \citet{2019ApJ...886..139N, 2020PhRvR...2a2046M}, which is written as
\begin{equation}
\sigma = \sqrt{-\left(\int_{G(\Omega_\nu)>0} \frac{d \Omega_\nu}{4\pi}G(\Omega_\nu)\right)\left(\int_{G(\Omega_\nu)<0} \frac{d \Omega_\nu}{4\pi}G(\Omega_\nu)\right)},
\end{equation}
where $\Omega_\nu = (\theta_\nu,\,\phi_\nu)$ is the neutrino flight direction. In the expression, $G(\Omega_\nu)$ denotes ELN angular distribution, which is defined as
\begin{equation}
G(\Omega_\nu) =\sqrt{2}G_{\rm F} \int_{0}^\infty\frac{\epsilon^2 d\epsilon}{2\pi^2}(f_{\nue} - f_{\nueb}),
\end{equation}
where $G_{\rm F}$ denotes the Fermi constant, and $f_{\nue/\nueb}$ is the distribution function of $\nue/\nueb$.

\section{Supernova Properties} \label{sec:evolution}
\begin{figure}
    \includegraphics[width=\hsize,bb=0.000000 0.000000 328.000000 626.000000]{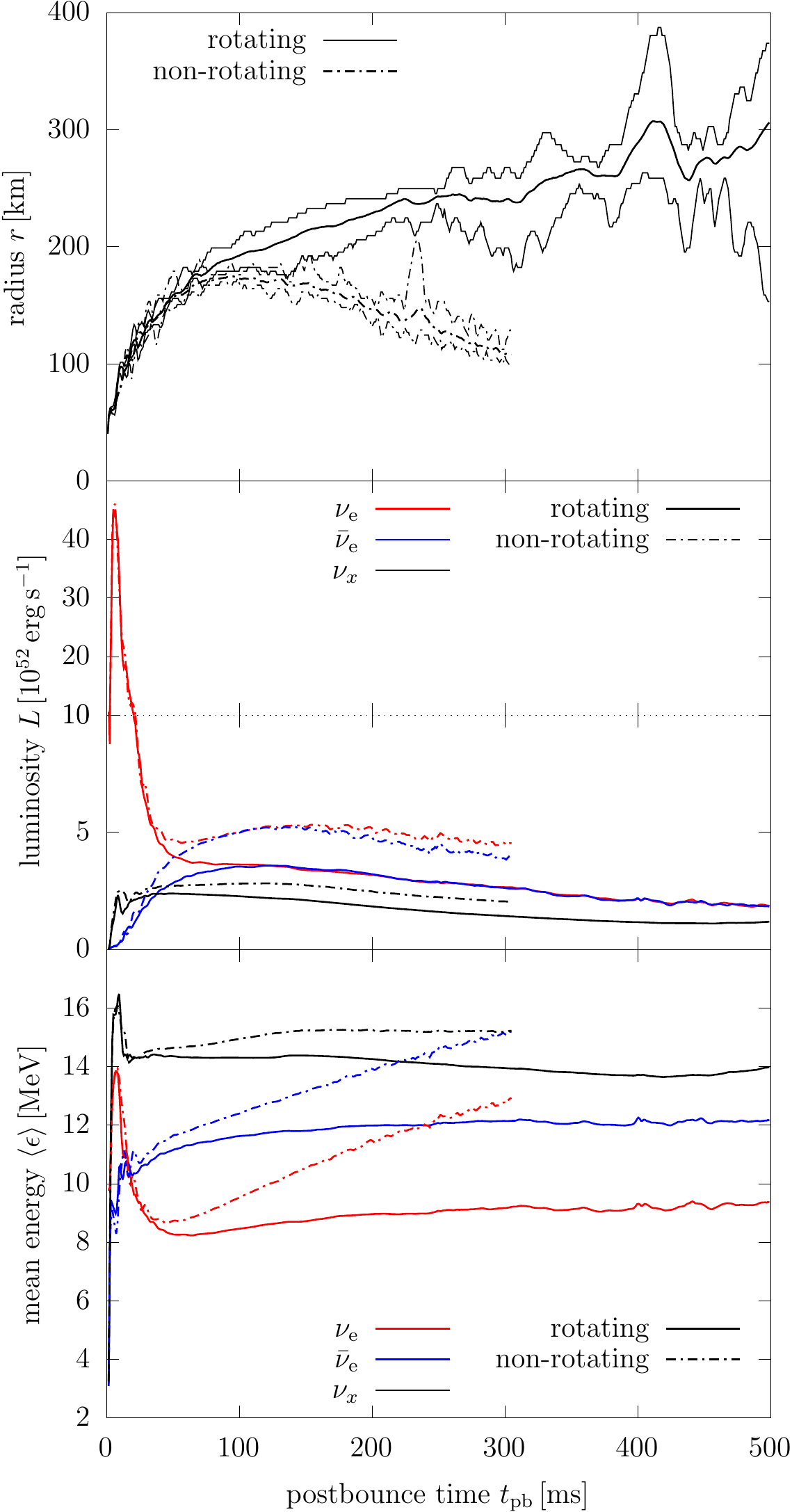}
    \caption{The time evolution of the shock radii, neutrino luminosities, and mean energies. The top panel shows the shock radii. The bold and thin lines represent the angular average and maximum/minimum shock radii. The middle panel indicates the neutrino luminosities measured at $r=500\,{\rm km}$. The red, blue, and black colors stand for $\nue$, $\nueb$, and $\nux$, respectively. The upper and lower halves, separated by the dotted line, have different perpendicular scales. The bottom panel indicates the neutrino mean energies, again measured at $r=500\,{\rm km}$. The colors are the same as the middle panel. For all panels, the solid and dash-dotted lines represent the rotating and non-rotating models, respectively.}
    \label{fig:timeevolution}
\end{figure}

Before entering details of the analysis of FFC, we describe overall properties of our CCSN models to catch the foundation on which we will give the subsequent discussion. Figure~\ref{fig:timeevolution} shows the time trajectory of shock radii, neutrino luminosities, and mean energies for both the rotating and non-rotating models. In the panel displaying the neutrino luminosity, we change the perpendicular scales for the upper and lower halves in order to see the neutronization burst and subsequent evolution of each flavor at once.

The shock radii in the rotating model expand further than the non-rotating counterpart. The non-rotating model shows a gradual decrease of the shock radii after reaching its maximum, failing in the shock revival. The rotating model shows continuous shock expansion. However, the shock evolution is quite gradual, and the average value reaches $\sim 300\,{\rm km}$ at the latest time. Hence, whether the supernova successfully explodes or not is still uncertain. More time is required for judging the fate of the dynamics. We are currently running the long-term simulation for the rotating model, and the results will be reported in another paper.

The neutrino luminosities for the rotating model are smaller than the non-rotating counterpart. This trend, the non-exploding model has high neutrino luminosities, is commonly observed in the supernova simulations \citep[see, e.g.,][]{2018ApJ...852...28S, 2018MNRAS.477.3091V}. Correspondingly, the neutrino mean energies (the angular average of the ratio of the neutrino energy density to the number density) for the rotating model is also smaller than the non-rotating model.

The $\nue$ and $\nueb$ luminosities are similar for both rotating and non-rotating models, while $\nue$ has lower mean energy than $\nueb$, indicating that the $\nue$ numbers are larger than $\nueb$ numbers. Furthermore, the difference in the $\nue$ and $\nueb$ mean energies are smaller for the non-rotating model than the rotating model, and hence the rotating model is more $\nue$ rich than the non-rotating model. This condition may be considered unfavorable for the FFC in the rotating model. However, it is compensated by the angular-dependent variation in the rotating model, as will be discussed in the following section.

\section{Fast Flavor Conversion} \label{sec:results}
\begin{figure}
    \includegraphics[width=\hsize,bb=0.000000 0.000000 307.000000 250.000000]{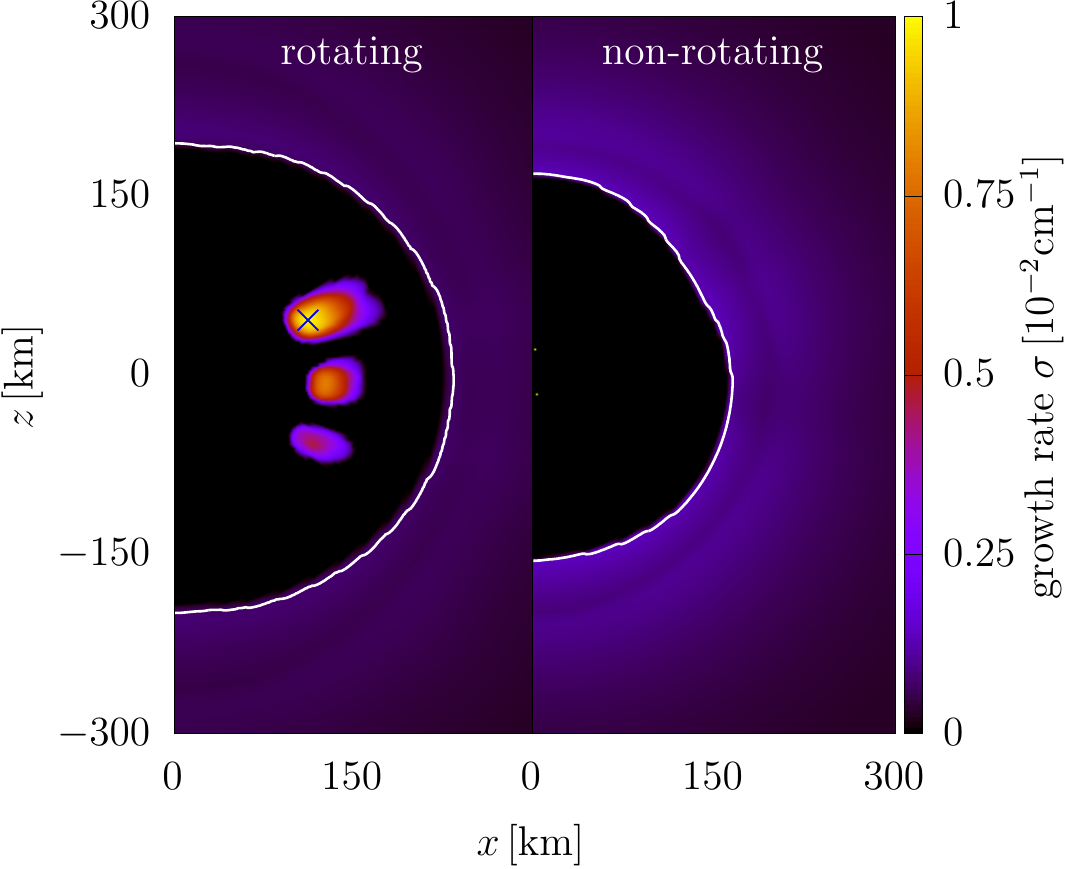}
    \caption{Spatial distributions of the linear growth rate of the flavor instability at $165\,{\rm ms}$ after the core bounce for the rotating (left panel) and non-rotating (right panel) models. White curves indicate the shock radii. The blue cross is the point discussed in Figure~\ref{fig:ELNcross}.}
    \label{fig:growth}
\end{figure}

As shown in the left panel of Figure~\ref{fig:growth}, the ELN crossing appears in the equatorial region for the rotating model, and the resultant growth rate of FFC seems to be fast ($\mathcal{O}(10^{-2})\,{\rm cm}^{-1}$). It should be mentioned that we have never seen this trend before in the previous studies; ELN crossings appear only in the equatorial region \citep[see, e.g.,][]{2021arXiv210807281N}. In fact, these crossings are not observed in the non-rotating model (see the right panel of Figure~\ref{fig:growth}). These facts suggest that the stellar rotation has some roles on the trend. Providing its physical interpretation is the subject in this paper\footnote{ELN crossings commonly appear in the pre-shock region for both rotating and non-rotating models. This comes from the disparity of inward-scattering by heavy nuclei between $\nue$ and $\nueb$ \citep[see][in more detail]{2020PhRvR...2a2046M}.}.

The absence of the FFC in the post-shock region of the non-rotating model is in line with the previous studies. \citet{2021arXiv210807281N} conducted a systematic study of occurrences of the FFC in the non-rotating CCSNe models. They investigated many 3D CCSNe simulations with moment-scheme neutrino transport. They reported that the FFC occurs at $\sim 100\,{\rm ms}$ after the core bounce for all supernova models and after that mainly for exploding models. They also discussed the possibility of the FFC by the PNS convection, but it is hardly seen because of the limited spatial resolution \citep{2020PhRvD.101f3001G}. Their study is consistent with other ELN crossing searches in CCSN simulations with the full neutrino distribution functions \citep{2019ApJ...886..139N, 2019PhRvD..99j3011D, 2020PhRvD.101b3018D, 2019PhRvD.100d3004A, 2020PhRvD.101d3016A} and approximate crossing searches \citep{2020JCAP...05..027A, 2021PhRvD.103f3033A}. Considering the fact that the non-rotating model in our model fails to explode (see Section~\ref{sec:evolution}), no FFC at $165\,{\rm ms}$ is compatible with these previous works.

\begin{figure}
    \includegraphics[width=\hsize,bb=0.000000 0.000000 279.000000 130.000000]{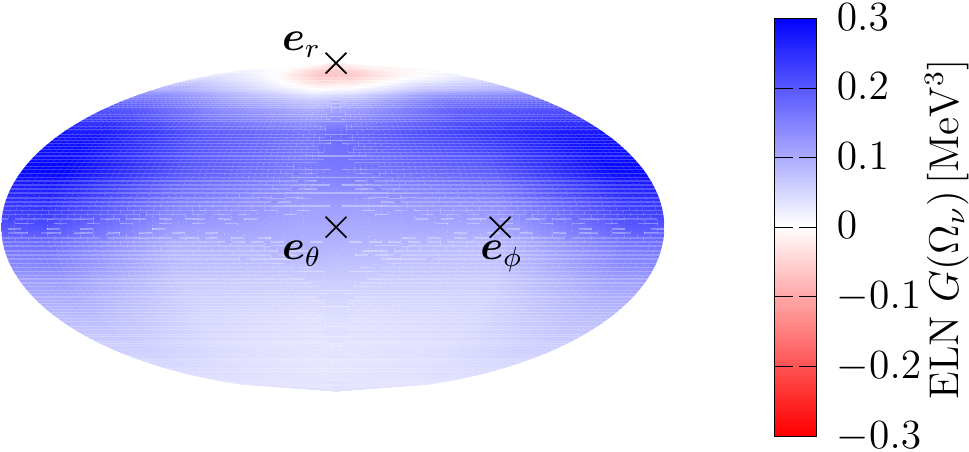}
    \caption{The Mollweide projection of the angular distribution of the ELN at the blue cross in Figure~\ref{fig:growth}. The crosses with $\bs e_r$, $\bs e_\theta$, and $\bs e_\phi$ represent $\theta_\nu=0$ (radially outgoing direction), $(\theta_\nu, \phi_\nu)=(\pi/2,0)$ (perpendicular to the radial direction, but inside the meridian plain), and $(\theta_\nu, \phi_\nu)=(\pi/2,\pi/2)$ (direction of the rotation), respectively. Bluish and reddish colors indicate the $\nue$ and $\nueb$ excesses in the direction, respectively.}
    \label{fig:ELNcross}
\end{figure}

In order to see the origin of the ELN crossing, let us first take a look at the angular distribution of ELN for the rotating model. The representative example ($r=120\,{\rm km}$ and $\theta=1.18\,{\rm rad}$) is displayed in Figure~\ref{fig:ELNcross}. There are more $\nueb$ in the outgoing direction than $\nue$, while vice versa for the other directions. Following the language used in \cite{2021arXiv210807281N}, this is type-II ELN crossing. We note that, although the stellar rotation directs the neutrino flux to the rotational direction \citep{2019ApJ...872..181H}, it is not strong enough to have an influence on the ELN crossing.

\begin{figure}
    \includegraphics[width=\hsize,bb=0.000000 0.000000 300.000000 583.000000]{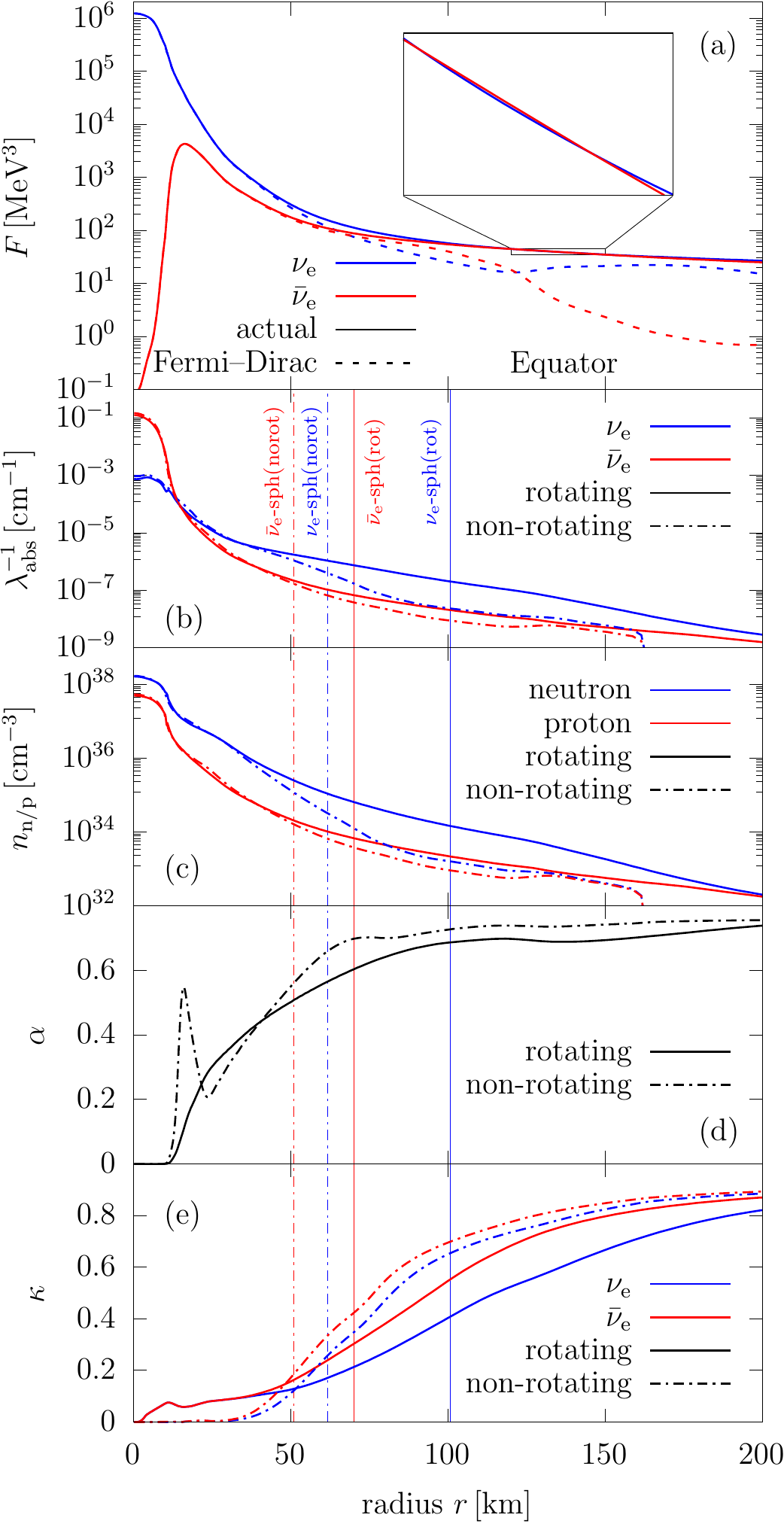}
    \caption{The radial profiles of the physical quantities relevant for the FFC along the equator. Panel (a): the energy-integrated distribution function $F$ for the radially flying ($\theta_\nu=0$) neutrinos. Blue and red colors correspond to $\nue$ and $\nueb$, respectively. The solid and dashed lines represent $F_{\nue/\nueb}$ and $F_{\nue/\nueb}^{\rm FD}$, respectively. The inset is the zoom-in figure of the indicated area, where the ELN crossing takes place. Panel (b): the absorptivity, the inverse of the absorption mean free path, for neutrinos with $\epsilon = 11\,{\rm MeV}$. The blue (red) color indicates the absorptivity for the (anti)neutrinos. Panel (c): the number density of the neutrons and protons. The blue stands for the neutrons, and the red is for the protons. Panel (d): $\alpha$ defined in the text. Panel (e): the flux factor. The blue and red colors are for $\nue$ and $\nueb$, respectively. From panels (b) to (e), we highlight the neutrinosphere for $\nue$ (blue) and $\nueb$ (red) by vertical lines. The line type distinguishes the rotating (solid) and non-rotating (dashed-dotted) models.}
    \label{fig:EWprofiles}
\end{figure}

Panel (a) of Figure~\ref{fig:EWprofiles} shows the radial profiles of the energy-integrated distribution function of neutrinos in the outgoing direction ($F_{\nue/\nueb} := \int_0^\infty f_{\nue/\nueb} \epsilon^2 d\epsilon$), and their Fermi--Dirac (FD) distributions determined locally from thermal and chemical equilibrium with matter ($F_{\nue/\nueb}^{\rm FD} := \int_0^\infty f_{\nue/\nueb}^{\rm FD} \epsilon^2 d\epsilon$, where $f_{\nue/\nueb}^{\rm FD}$ denotes the FD distribution function for $\nue/\nueb$). We select these profiles on the equator in the rotating model. As shown in the panel, although $\nue$ is more populated than $\nueb$ at $\la 100\,{\rm km}$, it decreases sharply with radius. Eventually, $\nueb$ dominates over $\nue$ at $\sim 150\,{\rm km}$ (albeit tiny). The increase or decrease of neutrinos should be dictated by neutrino--matter interactions; hence, we look into the collision term in detail. We note that nucleon scatterings play negligible roles in generating the crossings (we confirmed); thus, we focus only on emission and absorption processes hereafter.

Both $F_{\nue}$ and $F_{\nueb}$ coincide with $F_{\nue}^{\rm FD}$ and $F_{\nueb}^{\rm FD}$, respectively, at the central region (optically thick region). In the outer regions, on the other hand, $F^{\rm FD}$ becomes smaller than $F$ regardless of neutrino species. This indicates that neutrino absorption dominates the emission, considering that the collision term (negative means absorption) is expressed as
\begin{equation}
S_{\rm rad,\,\nue/\nueb} = -\frac{1}{\lambda_{\rm abs,\,\nue/\nueb}}(f_{\nue/\nueb}-f_{\nue/\nueb}^{\rm FD}), \label{eq:collision}
\end{equation}
where $\lambda_{\rm abs,\,\nue/\nueb}$ is the mean free path for the absorption of $\nue/\nueb$. As shown in panel (b) of Figure~\ref{fig:EWprofiles}, the absorptivity of $\nue$ is higher than $\nueb$, and more importantly, the difference is much larger than the case with the non-rotating model. Here, we define the absorptivity as the inverse of $\lambda_{\rm abs,\,\nue/\nueb}$ and display them for $11\,{\rm MeV}$ neutrinos. The different absorptivity between $\nue$ and $\nueb$ reflects the disparity of the number density of interacting nucleons \citep[see also][]{2021arXiv210807281N}. As displayed in panel (c) of Figure~\ref{fig:EWprofiles}, the neutron number density is much higher than proton; consequently, $\nue$ experiences strong absorption. These facts suggest that the disparity between $\nue$- and $\nueb$- absorption is responsible for the ELN crossing. We also note that the $F_{\nueb}^{\rm FD}$ is larger than $F_{\nue}^{\rm FD}$ at $60\,{\rm km} \la r \la 120\,{\rm km}$ (see panel (a) of Figure~\ref{fig:EWprofiles}), indicating that the chemical potential of $\nue$ becomes negative in this region. In such a region, $\nue$ absorption is much stronger than $\nueb$, which facilitates the occurrence of ELN crossing.

We now turn our attention to the effect of rotation. The change of matter distribution by the centrifugal force is the key. In general, the density gradient in the equatorial region becomes shallower than the non-rotating cases (due to the centrifugal force), indicating that the low-$\ye$ (electron fraction) environment in the post-shock region is extended. This provides a preferable condition for the excess of neutrons to protons, which is why $\nue$ goes through strong absorption.

It is interesting to be pointed out that $\alpha$ (the ratio of the number density of $\nueb$ to $\nue$) in the region with the ELN crossing is $\sim 0.7$ (see panel (d) of Figure~\ref{fig:EWprofiles}), which is smaller than that has been observed in non-rotating CCSN models \citep[see, e.g.,][]{2019PhRvD.100d3004A, 2020PhRvD.101b3018D, 2019ApJ...886..139N, 2021arXiv210807281N}. Instead, the differences in the degree of forward-peaking angular distributions between $\nue$ and $\nueb$ seem large enough to make the ELN crossings. To quantify the degree of forward-peaking, we compute the flux factor ($\kappa$, defined as the number flux divided by the number density) of $\nue$ and $\nueb$ for both the rotating and non-rotating models\footnote{We note that $\kappa$ is a useful quantity to characterize the full angular distributions of neutrinos in CCSNe; see \citet{2021arXiv210405729N} for more detail.}, and the results are displayed in the panel (e) of Figure~\ref{fig:EWprofiles}. As shown in the panel, the difference in the flux factors between $\nue$ and $\nueb$ for the rotating model is much larger than the non-rotating model. This suggests that $\nueb$ has a much sharper forward-peaked angular distribution than $\nue$ in the rotating model; hence, $\nueb$ in the outgoing direction dominates over $\nue$ (in spite of low $\alpha$).

The large difference in the angular distributions between $\nue$ and $\nueb$ can be understood by considering the positions of neutrinospheres. From panels (b) to (e) of Figure~\ref{fig:EWprofiles}, we highlight the positions of neutrinospheres as vertical lines (for comparison, those in the non-rotating model are also displayed). As shown in these panels, the neutrinosphere radii for both $\nue$ and $\nueb$ tend to be larger in the rotating model than in the non-rotating model. This is due to the shallow density structure in the equatorial region of the rotating model. It should also be mentioned that the low-$\ye$ environment in the equatorial region increases the difference in the neutrinosphere radii between $\nue$ and $\nueb$ since the large $\nue$ absorption rate increases the neutrinosphere radius of $\nue$. As a result, the degree of forward-peak in $\nue$ angular distribution is substantially reduced.

\begin{figure}
    \includegraphics[width=\hsize,bb=0.000000 0.000000 195.000000 213.000000]{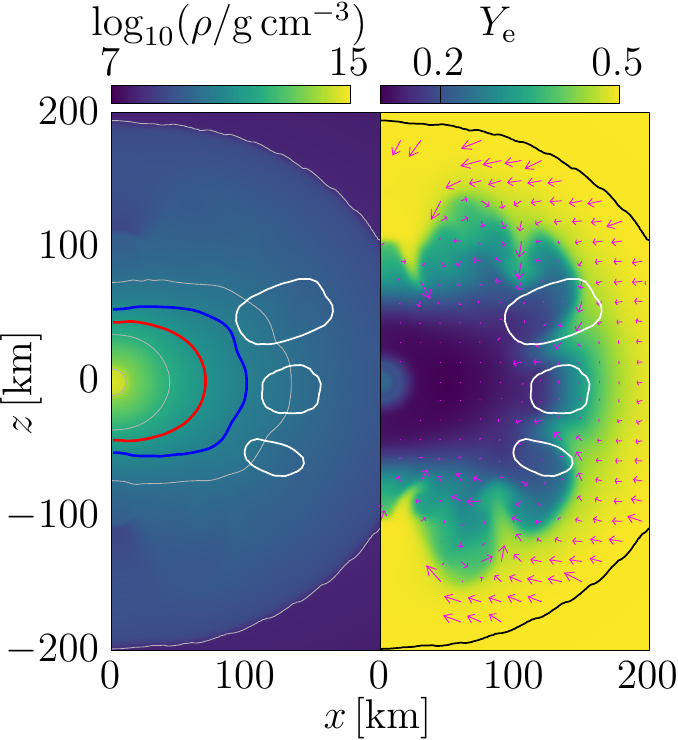}
    \caption{The color map of the multi-D hydrodynamic configurations. Left: density $\rho$ distribution for the rotating model. The gray curves show the contours with the density of $10^{14}$, $10^{12}$, $10^{10}$, and $10^{8}\,{\rm g\,cm^{-3}}$ from inside to outside. The blue and red curves are the neutrinospheres for $\nue$ and $\nueb$, respectively, whose energies are $11\,{\rm MeV}$. Right: $\ye$ for the rotating model. The magenta arrows represent the fluid velocity field, with speed being proportional to the length of the arrow. The shock radius is denoted by the black curve. For both panels, the enclosed regions with white curves indicate the flavor-unstable regions.}
    \label{fig:map_den_ye}
\end{figure}

Here, we discuss the latitudinal dependence of ELN crossing. In the left panel of Figure~\ref{fig:map_den_ye}, we portray the neutrinospheres on the top of the matter-density color map. This illustrates that the neutrinospheres are oblately deformed, and more importantly, the neutrinospheres of $\nue$ and $\nueb$ tend to be close around the polar regions, which is attributed to the sharp decline of the matter density and absorptivity. Consequently, the difference of angular distributions between $\nue$ and $\nueb$ becomes small, which suppresses the occurrence of ELN crossings around polar regions. We also note that there is inhomogeneity of ELN crossing in the equatorial region. We first point out that the origin is associated with convective fluid motions. Since the neutrino absorption dominates over emission at $\ga 100\,{\rm km}$, the negative entropy gradient in the radial direction is sustained, which drives convection (i.e., neutrino-driven convection). Due to the convective motion, high-$\ye$ clumps sink into the smaller radius, and low-$\ye$ buoyant plums propagate outward (see the right panel of Figure~\ref{fig:map_den_ye}). In the high-$\ye$ clumps, the disparity of number density between neutron and proton becomes small; consequently, the difference in $\nue$ and $\nueb$ absorption rates is reduced, which suppresses the occurrence of ELN crossings.

\section{Summary and Conclusion} \label{sec:concl}
ELN crossings emerge around the post-shock equatorial region in our rotating model, whereas they are not observed in the non-rotating model. We confirm that the crossings are induced by rotation; the essential physical mechanism is as follows. Centrifugal force deforms the matter distribution oblately in the post-shock region. As a result, the low-$\ye$ region is extended to a larger radius than the non-rotating model. Since the neutrons are more abundant than the protons in the low-$\ye$ region, the outgoing $\nue$ experiences more absorption by matter than $\nueb$, which facilitates the generation of the crossings. We also find that $\alpha$ in the ELN crossing region is $\sim 0.7$, which is smaller than reported in other non-rotating CCSN models \citep{2019PhRvD.100d3004A, 2020PhRvD.101b3018D, 2019ApJ...886..139N, 2021arXiv210807281N}. This suggests that the ELN crossing reflects the large difference of angular distributions between $\nue$ and $\nueb$. The rotation is responsible for the difference; the two conditions, shallow density gradient and low-$\ye$ environment, push the neutrinosphere of $\nue$ outward; consequently, $\nueb$ has a more forward-peaked angular profile than $\nue$, leading to ELN crossings. In short, $\nue$ is absorbed by low-$\ye$ matter extended by centrifugal force to be more isotropic than $\nueb$. Finally, we inspect the latitudinal dependence of the ELN crossing and find that neutrino-driven convection primarily accounts for the inhomogeneity of the appearance of ELN crossings around the equatorial region.

Since neutrinos are associated with many ingredients of CCSN, the observables such as neutrino signals, morphology of explosions and nucleosynthesis may be substantially impacted by the rotation-induced FFC. Our finding of this paper opens a new window for the connection between FFC and CCSN theory. For more general arguments, we need a systematic study of FFC in rotating CCSNe by changing rotations, progenitors, and their time dependence; the detailed investigations are postponed to future work.

\section*{Acknowledgments}
%\acknowledgments % To editor: the \acknowledgments command does not work well in my tex environment, hence we use \section command instead.
We thank T. Morinaga, M. Zaizen, S. Yamada, W. Iwakami, and T. Takiwaki for fruitful discussions. This research used high-performance computing resources of K-computer by R-CCS, FX10 and Oakforest-PACS by the University of Tokyo, Grand Chariot by Hokkaido University, and FX100 by Nagoya University, through the HPCI System Research Project (Project ID: hp 140211, 150225, 160071, 160211, 170230, 170031, 170304, 180111, 180239, 190100, 200102), SR16000 and XC40 at YITP of Kyoto University, SR16000 and Blue Gene/Q at KEK under the support of its Large Scale Simulation Program (14/15-17, 15/16-08, 16/17-11), Research Center for Nuclear Physics (RCNP) at Osaka University, and the XC30 and the general common-use computer system at the Center for Computational Astrophysics, CfCA, the National Astronomical Observatory of Japan.  This work was supported in part by a Grant-in-Aid for Scientific Research on Innovative areas “Gravitational wave physics and astronomy: Genesis” (17H06357, 17H06365) from the Ministry of Education, Culture, Sports, Science and Technology (MEXT), Japan, and in part a Grant-in-Aid for Scientific Research (C; 15K05093, 19K03837, B; 20H01905) and Young Scientists (Start-up, JP19K23435) from the Japan Society for the Promotion of Science (JSPS). This work was also partly supported by research programs at K-computer of the RIKEN R-CCS, HPCI Strategic Program of Japanese MEXT, Priority Issue on Post-K-computer (Elucidation of the Fundamental Laws and Evolution of the Universe), Joint Institute for Computational Fundamental Sciences (JICFus), and by MEXT as “Program for Promoting Researches on the Supercomputer Fugaku” (Toward a unified view of the universe: from large scale structures to planets).

%\vspace{5mm}
\software{Gnuplot \citep{gnuplot5}}



%\appendix

%\section{Appendix information}
%possible appendix


\bibliography{ref}%{}
\bibliographystyle{aasjournal}



\end{document}